%% file: feedback_itw.tex
\documentclass[10pt,conference]{IEEEtran}
\pdfoutput=1
\usepackage{layout, calc}
\usepackage{setspace}
\usepackage{times,amssymb,amsmath,epsfig,amsthm}
\usepackage{citesort}
\usepackage{subfigure}
 \DeclareGraphicsExtensions{.pdf,.jpeg,.png,.pdf_t}
\usepackage{color}

\newtheorem{theorem}{Theorem}

   \newtheorem{remark}{Remark}

\title{On Channel Output Feedback in Deterministic Interference Channels}
\author{
\authorblockN{Achaleshwar Sahai}
\authorblockA{
Department of ECE,\\
 Rice University,\\
Houston, TX 77005.\\
as27@rice.edu}\\
\and
\authorblockN{Vaneet Aggarwal}
\authorblockA{
Department of ELE,\\
Princeton University,\\
Princeton, NJ 08544.\\
vaggarwa@princeton.edu}\\
\and
\authorblockN{Melda Y\"{u}ksel}
\authorblockA{
Department of EEE,\\
TOBB University of \\Economics and Technology,\\
Ankara, Turkey.\\
    yuksel@etu.edu.tr}\\
 \and
\authorblockN{Ashutosh~Sabharwal}
\authorblockA{
Department of ECE,\\
 Rice University,\\
Houston, TX 77005.\\
ashu@rice.edu}\\
 }
\date{}
\begin{document}
\maketitle
\begin{abstract}
In this paper, we study the effect of channel output feedback on the sum capacity in a two-user symmetric deterministic interference channel. We find that having a single feedback link from one of the receivers to its own transmitter results in the same sum capacity as having a total of 4 feedback links from both the receivers to both the transmitters. Hence, from the sum capacity point of view, the three additional feedback links are not helpful. We also consider a half-duplex feedback model, where the forward and the feedback resources are symmetric and time-shared. Surprisingly, we find that there is no gain in sum-capacity with  feedback in a half-duplex feedback model, when interference links have more capacity than direct links.
\end{abstract}

\begin{section}{Introduction}
It is well known that for point-to-point channels, feedback does not improve the capacity of discrete-memoryless channels \cite{shannon56}. Feedback in multiple-access channels (MAC) does enlarge the achievable rate region \cite{gwolf,ozzarow} but does not provide multiplexing gain. However, unlike MAC, it was shown in \cite{suh-2009} that feedback increases multiplexing gain in a deterministic interference channel, when there are two dedicated feedback links in the system, each from one of the receivers to its own transmitter. 

The deterministic model for interference channel was introduced in \cite{costa82}. The model in \cite{costa82} was fairly general and was specialized as an approximation to the linear Gaussian model without feedback in \cite{atdintro}. The deterministic model was further extended to interference channels in \cite{bresler2,jafar} without feedback, where the authors use a model which is a special case of the model proposed for interference channel. Channel output feedback in interference channels has been extensively studied in \cite{suh-2009,kramer02,kramer04,gast06,jiang07}. 


In this paper, we study three different feedback models. In the first model, only one receiver can send feedback to its transmitter. This models an extreme case of asymmetry in feedback, which can possibly exist in systems providing different quality of service to different users. We study this asymmetry by denying one user of all feedback, while providing full feedback to the other user. In the second model, both the receivers can send feedback to both the transmitters, resulting in a total of four feedback links. The feedback from each of the receivers could potentially be heard by both the transmitters and can be used by them to make superior decisions in order to improve the sum-capacity.
We find that the sum capacity in these two variants (one and four links of feedback) is the same as that in \cite{suh-2009}. Thus, with respect to sum-capacity, performance of one feedback link is  as good as two or four feedback links. The intuition behind this result is that the single feedback link can aid cooperation in such a way that the common rate of one of the users (user without feedback) can now be doubled by sacrficing the common rate of cooperating user. In the third model we study a practical time-division duplex (TDD) system, where there are no dedicated feedback links. The feedback is broadcast on the channel with same link capacities between any two nodes as that in the forward channel. Moreover, time (resource) is shared among forward and feedback channels and a loss in rate is incurred. We model the above constraints as a half-duplex interference channel and find that feedback does not increase the sum capacity in a strong interference channel. The receivers receive data from two independent paths, first being the direct link and the second being the path involving feedback link of the other user. The link capacity of the feedback being same as the forward link acts as a bottleneck for total data being received through the second path. Thus presence of an interference link with high link capacity does not help increase the sum-capacity in strong interference regime, unlike \cite{suh-2009}, in which the authors considered dedicated feedback links.

The rest of the paper is organized as follows. In Section II, we give the channel model. In Section III, we find the sum-capacity of a feedback scheme when only one of the transmitters receives feedback from its receiver. In Section IV, we add dedicated cross feedback links and find the sum-capacity. In either case sum-capacity is found to be the same. Section V elaborates the half-duplex feedback model. We conclude the paper  with some remarks and intuitions in Section VI.

\begin{figure}[htb]
 \begin{center}

  \resizebox{1.8in}{!}{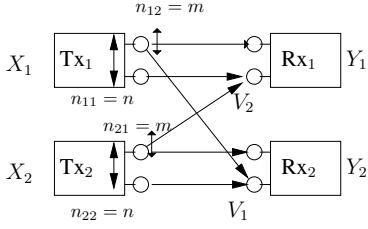}
   \caption {Symmetric deterministic interference channel}
  \label{fig:model}
  \end{center}
\end{figure}

\end{section}

\begin{section}{Channel Model and Preliminaries}
A two-user interference channel consists of two point-to-point transmitter-receiver links, where
the receiver of each link also receives an interfering signal from the
unintended transmitter. In a deterministic interference channel, the input $X_{ki}$ to  the  $k^{\mathrm{th}}$
transmitter at time  $i$ can be written as $X_{ki}=\left[
X_{ki_1}\text{ }X_{ki_2}\text{ }\ldots X_{ki_q}\right]^T  $, $k=1,2,$  such that $X_{ki_1}$ and
$X_{ki_q}$ are the most and the least significant bits, respectively.
The received signal of user $j$, $j=1,2$, at time $i$  is denoted by the vector
$Y_{ji}=\left[  Y_{ji_1}\text{ }Y_{ji_2}\text{ }\ldots\text{ }Y_{ji_q}\right]
^{T}$.
Associated with each transmitter $k$ and receiver $j$ is a non-negative
integer $n_{kj}$ that defines the number of bit levels of $X_{ki}$ observed at receiver $j$, at time $i$. The maximum level supported by any link is $q=\max_{j,k}(n_{jk})$.
Specifically, the received signal
$Y_{ji}$, $j=1,2,$ of an interference channel is given by%
\begin{equation}%
\begin{array}
[c]{cc}%
Y_{ji}=\mathbf{S}^{q-n_{1j}}X_{1i}\oplus\mathbf{S}^{q-n_{2j}}X_{2i} &
j=1,2,
\end{array}
\end{equation}
where $\oplus$ denotes the XOR operation, and $\mathbf{S}^{q-n_{jk}}$ is a $q\times q$ shift matrix with entries
$S_{m,n}$ that are non-zero only for $\left(  m,n\right)  =(q-n_{jk}%
+n,n),n=1,2,\ldots,n_{jk}$.

In this paper, we consider a symmetric deterministic interference channel as shown in Fig. \ref{fig:model}. The symmetric channel is characterized by two values: $n = n_{11} = n_{22}$ and $m = n_{12} = n_{21}$, where $n$ and $m$ indicate the number of signal bit levels that we can send through direct links and the cross links, respectively.

There are two independent and uniformly distributed sources, $W_k \in \{1, 2, \cdots, M_k\}, \forall k =
1, 2$, where $M_{k}$ is total number of codewords of the $k^{\text{th}}$ user. Due to feedback, the encoded signal $X_{ki}$ is a function of its own message $W_{k}$ and past output sequences $Y_{k}^{i -1}$. The sum capacity is defined as
\begin{eqnarray}
C_{sum} = \sup \{R_1+R_2 : (R_1,R_2) \in {\cal R}\} ,
\end{eqnarray}
where ${\cal R}$ is the capacity region. In this paper, we consider four models of feedback as described below.
\begin{enumerate}
\item Two-link feedback model: This model has a dedicated feedback link from each of the receiver to its own transmitter. This model has been studied in \cite{suh-2009}. In this model, the encoded signal $X_{ki}$ of user $k$ at time $i$ is a function of its own message and past output sequences of its own receiver. Thus,
    \begin{equation}
    X_{ki}=f_k^i(W_k,Y_k^{i-1}),
    \end{equation}
    where $Y_k^{i-1}$ is shorthand notation for $Y_{k1}, \cdots, Y_{k i-1}$.
\item One-link feedback model: In this model, there is a dedicated feedback link from the first receiver to the first transmitter as shown in Fig. \ref{fig:onesided_model}. However, the second transmitter does not receive any feedback. Hence, although the first transmitter encoding scheme depends on $W_1$ as well as the past outputs, the encoding scheme of the second transmitter is only a function of $W_2$.

\begin{figure}[htb]
 \begin{center}

  \resizebox{1.8in}{!}{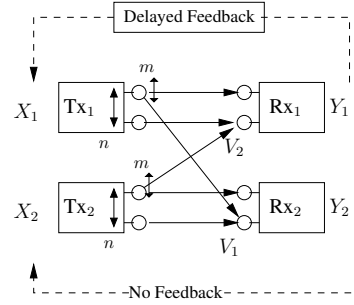}
   \caption {Single Feedback Link}
  \label{fig:onesided_model}
  \end{center}
\end{figure}

\item Four-link feedback model: In this model, we consider dedicated feedback link from each receiver to both the transmitters as shown in Fig. \ref{fig:fourlink}. Hence, the encoding strategies at each of the transmitter is a function of its own message and the past output sequences of both the receivers, or
    \begin{equation}
    X_{ki}=f_k^i(W_k,Y_1^{i-1},Y_2^{i-1}).
    \end{equation}

\begin{figure}[htb]
 \begin{center}

  \resizebox{1.8in}{!}{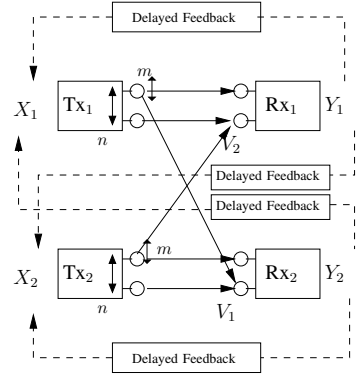}
   \caption {Four Feedback Links}
  \label{fig:fourlink}
  \end{center}
\end{figure}
\item Half-duplex feedback model: In this model, the feedback link is also an interference channel, as shown in Fig.\ref{fig:reci}. Let the two tuple $[n,m]$ be defined as the $\emph{operating-point}$ of the symmetric deterministic channel, where $n$ and $m$ indicate the number of signal bit levels that we can send through the direct links and the cross links respectively. We study the case, where the operating point of both the forward as well as the feedback channel are the same, i.e., the link capacities in the feedback channel are the same as in the forward channel. Further, the use of the channels are time divided in the sense that the forward channel is used for fraction $t$ of the time while the feedback channel is used $1-t$ fraction of the time, where $t \in [0,1]$ is a parameter that can be controlled.
\end{enumerate}

The first case with two feedback links from each receiver to its own transmitter was considered in \cite{suh-2009}, where the sum capacity in the presence of feedback was analyzed.
\begin{theorem}[\cite{suh-2009}]\label{2sided} The feedback sum capacity of a deterministic interference channel, when there are dedicated feedback links from each receiver to its own transmitter, is given by
\begin{eqnarray}
C_\text{sum}= \max(n,m)+(n-m)^+.
\end{eqnarray}
\end{theorem}

%
%

\end{section}

\begin{section}{One-link Feedback model}
In this section, we explore the case where only one of the users is allowed full feedback, while the other user is devoid of any kind of feedback.

\begin{theorem}\label{1side} The feedback sum capacity of a deterministic interference channel, when there is a single dedicated feedback link from the first receiver to its own transmitter, is given by
\begin{eqnarray}
C_{\text{sum}}= \max(n,m)+(n-m)^+.
\end{eqnarray}
\end{theorem}
\begin{proof}
The proof is provided in Appendix \ref{proof1side}. The proof uses the fact that data can travel from the second transmitter to second receiver through two paths now. The first is the direct path and the second is the path through the second transmitter, the first receiver, the first transmitter to the second receiver. 
In case of very strong interference ($\frac{m}{n} \geq 2$), we show the achievability of the outer bound which is $m$. Consider the scenario, where the first transmitter does not send any data to its receiver. The transmitter-receiver pair along with feedback can then form a virtual node and act as a relay. It is easy to see this simplification in Fig. \ref{fig:relay}. The link via relay itself has a capacity of $m$ bits which is the outer bound (using Theorem 1). Thus, the sum-capacity itself is $m$. For weak interference ($\frac{m}{n} \le \frac{1}{2}$), the second transmitter sends i.i.d. data on all the $n$ bit locations. The first transmitter, on $(n-m)$ bit locations among all the interfering bit locations at receiver 2, transmits the data of the second user that interfered at the first receiver in the previous time slot as shown Fig. \ref{fig:one-sided feedback}. This way, the second receiver knows the interference and the first receiver can resolve the data (sequentially) from the interference. Hence, the rates $(R_1,R_2)=(n-m,n)$ can be achieved. The achievability for other regions is shown in  Appendix \ref{proof1side}.
\end{proof}
\begin{figure}[htb]
 \begin{center}

  \resizebox{1.7in}{!}{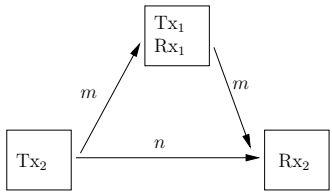}
   \caption {Transmitter Receiver pair acting as a virtual relay}
  \label{fig:relay}
  \end{center}
\end{figure}
\begin{remark}
Loss of one of the feedback links compared to the two feedback links considered in the channel model of Theorem \ref{2sided} does not decrease the sum capacity.
\end{remark}

\end{section}

\begin{section}{Four-link feedback model}
\label{4link}

In this section, we consider the case of interference channel with four feedback links.

\begin{theorem}
\label{th:fourlink} The feedback sum capacity of a deterministic interference channel, when there are dedicated feedback links from each receiver to both the transmitters, is given by
\begin{equation}
C_\text{sum}= {\max(n,m)+(n-m)^+}.
\end{equation}
\end{theorem}

\begin{proof} The achievability follows from \cite{suh-2009}. We will now prove the converse.

Let $W_{j}$ and $X_{j}$ be the message and transmitted vector at the $j^{\mathrm{th}}$ transmitter. Also, let $Y_{j}$ be received vector at the $j^{\mathrm{th}}$ receiver and $V_{j}$ be the interfering vector due to the $j^{\mathrm{th}}$ transmitter. The received vectors $Y_{1}$ and $Y_{2}$ are fed back to both the transmitters. We therefore have
\begin{eqnarray*}
&&N(R_1 + R_2)\nonumber\\ & \leq & H(W_1) + H(W_2) \\
& \stackrel{\text{(a)}}{=} &  H(W_1|W_2) + H(W_2) \\
& \stackrel{\text{(b)}}{\leq} & I(W_1; Y_1^N|W_2) + I(W_2;Y_2^N) + N\epsilon \\
& = & [H(Y_1^N|W_2) - H(Y_1^N|W_1,W_2) ] \nonumber\\&&+ [H(Y_2^N) - H(Y_2^N|W_2)] + N\epsilon  \\
& \stackrel{\text{(c)}}{=} & H(Y_1^N|W_2) - H(Y_2^N|W_2) + H(Y_2^N) + N\epsilon \\
& = & H(Y_1^N|W_2) +H(Y_1^N|Y_2^N,W_2) - H(Y_1^N,Y_2^N|W_2)\nonumber\\&& + H(Y_2^N) + N\epsilon\\
& = & H(Y_1^N|Y_2^N, W_2) - H(Y_2^N|Y_1^N,W_2) + H(Y_2^N) + N\epsilon \nonumber \\
& \leq & H(Y_1^N|Y_2^N,W_2)+ H(Y_2^N) + N\epsilon \\
& \stackrel{\text{(d)}}{=} & \sum_{i=1}^N H(Y_{1,i}| Y_{2}^N, Y_{1}^{i-1}, W_2) + H(Y_2^N) + N\epsilon \\
& \stackrel{\text{(e)}}{=} & \sum_{i=1}^N H(Y_{1,i}| Y_{2}^N, Y_{1}^{i-1}, W_2, X_{2}^i, V_2^i, V_1^{i}) + H(Y_2^N)+ N\epsilon \nonumber \\
& \stackrel{\text{(f)}}{\leq} & \sum_{i=1}^N H(Y_{1,i}| V_{1,i}, V_{2,i}) + H(Y_2^N) + N\epsilon ,
\end{eqnarray*}

\noindent where (a) follows from the independence of $W_1$ and $W_2$,  (b) follows by applying Fano's inequality to both the terms in (a), (c) follows from the fact that in a deterministic channel, the messages $W_1$ and $W_2$ together completely determine the output $Y^N_1$, (d) follows from chain rule of entropy, (e) follows from the observation that $X_{2}^i$ is a function of only ($W_2, Y_{1}^{i-1}, Y_{2}^{i}$), $V_2^{i}$ is function of $X_2^i$, and $V_1^{i}$ is a function of ($X_2^{i}$, $Y_{2}^i$), and (f) follows since conditioning reduces entropy.

By randomization of time indices and taking $\epsilon\to 0$ as $N \to \infty$, we get
\begin{equation}
R_1+R_2\le H(Y_1|V_1V_2)+H(Y_2).
\end{equation}
Since the RHS is maximized when $X_1$ and $X_2$ are uniform and independent, we get the converse as in the statement of Theorem \ref{th:fourlink}.
\end{proof}

\begin{remark}Two extra dedicated feedback links from each receiver to the other transmitter do not increase the sum capacity.
\end{remark}
\end{section}

\begin{section}{Half-duplex feedback model: Accounting for Feedback Resources}

For a TDD based system, in practice, the feedback channel shares the same resources as the forward channel. Hence, we consider the model in which each link has the same capacity in the forward and reverse directions. Both the forward and the feedback channels are interference channels and share the time resource between them. The forward channel can be used $t$ fraction of the time, and the feedback channel for the remaining $1-t$ fraction of the time. For $t=0$, no feedback is used while for $t=1$, no data is sent. In the following theorem, we prove  that the feedback does not help when the resources used in the feedback channel are accounted for, when $\frac{m}{n}\ge\frac{2}{3}$.

\begin{figure}[htb]
 \begin{center}

  \resizebox{1.6in}{!}{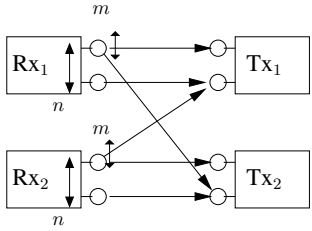}
   \caption {Feedback channel in Half duplex model}
  \label{fig:reci}
  \end{center}
\end{figure}

\begin{theorem}
Suppose that the forward and the feedback channels are both interference channels with the same operating point. Then, feedback does not increase the sum capacity for $m\ge\frac{2}{3}n$.
\end{theorem}
\begin{proof}
For the proof, we consider two separate ranges:
\begin{enumerate}
\item $\frac{2}{3}\le \frac{m}{n}\le 2$: Since any feedback strategy with dedicated feedback links (a maximum of 4) does not increase the sum capacity, accounting for the resources will not increase the sum capacity. Thus, using no feedback $\equiv$ $t = 1$ achieves the maximum rate.
\item $\frac{m}{n}\ge 2$: Consider an upper bound on the half-duplex model as shown in Fig.\ref{fig:hd1}, where all the links have been replaced by dedicated links. The fraction $t\in(0,1)$ indicates the fraction of time for which the forward link is in use. We further consider an upper bound of the system in Fig.\ref{fig:hd1} by letting the cross links ($\mathrm{Tx}_1\mathrm{-Rx}_2$ and $\mathrm{Tx}_2\mathrm{-Rx}_1$) to be of infinite capacity, as shown in Fig.\ref{fig:hd2}. The sum-total of the rate from the pair  $\mathrm{Tx}_1\mathrm{-Rx}_2$ to  $\mathrm{Tx}_2\mathrm{-Rx}_1$ and vice-versa is $n$. Since feedback in point to point channel does not increase capacity, therefore, the maximum rate at which transmission between $\mathrm{Tx}_1$ and $\mathrm{Rx}_1$ can happen is upper bounded by $n$.

    Hence, there is no improvement with feedback since the symmetric rate of $n$ can be achieved without feedback.

\end{enumerate}


 \begin{figure}[htb]
 \begin{center}
\subfigure[Half-Duplex Model with dedicated links]{

  \resizebox{1.3in}{!}{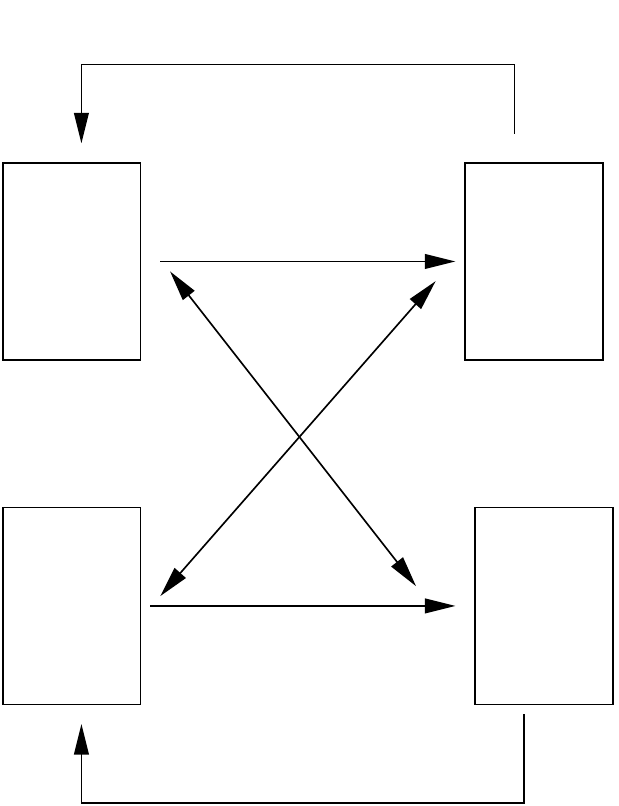}
\label{fig:hd1}
}
\subfigure[Upper bound on half-duplex model]{

  \resizebox{1.3in}{!}{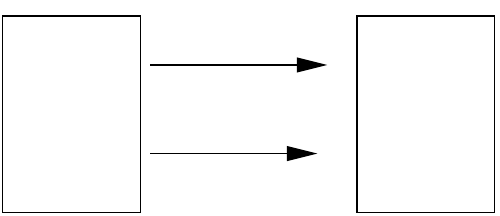}
\label{fig:hd2}
}
 \end{center}
  \caption {Two upper bounds on the half-duplex model}
 
\end{figure}
\end{proof}

\end{section}

\begin{section}{Discussion $\&$ Conclusions}
In this paper, we find that the feedback link from one receiver to its own transmitter gives the same sum capacity as the feedback link from both the receivers to both the transmitters for a symmetric deterministic interference channel. Hence, the three additional feedback links are not required from the sum capacity perspective. The one-link feedback forms a subset of all possible strategies over the four-link feedback scheme. It could therefore be possible that the one-link feedback may not be able to achieve all the points in the achievable rate region of the four-link feedback scheme.
These results extend to Gaussian channels with a constant bit gap between the one-link feedback model and four-link feedback model, and will be presented elsewhere.

Moreover, we find that when the feedback resources are symmetric and time shared with the forward channel, channel output feedback does not increase the sum capacity in a  deterministic interference channel in the strong interference regime. 

\end{section}

\appendices

\begin{section}{Proof of Theorem \ref{1side}}\label{proof1side}
The converse follows by adding one more feedback link from the second receiver to the second transmitter, and using Theorem \ref{2sided}. Hence, we focus on achievability in the proof.

We will frequently use $\max(n,n-m)$ and $\min(n,n-m)$ and for the sake of brevity will denote them by $l_{max}$ and $l_{min}$ respectively.

\noindent \emph{Case 1, $m/n\le 2/3$}:  Suppose $X_{1i}$ and $X_{2i}$ be the transmitted data from transmitter 1 and 2 at the $i^{\mathrm{th}}$ time slot.

\noindent \emph{Time slot 1:} The transmitted data vector at the $1^{\mathrm{st}}$ transmitter, $X_{1j}$ is comprised of $n$ i.i.d. bits. The data vector at transmitter 2 is such that the lower $m$ bits are set to 0 while the rest of the $(n -m)$ bits are independent of each other. The received vector $Y_{11}$ at receiver 1 therefore is
\begin{equation*}
  Y_{11_j} = \left\{
\begin{array}{ll}
X_{11_j} & \text{if }  1 \leq j \leq n -m \\
 X_{11_j}  \oplus X_{21_{j-n + m}} &  \text{if } n -m < j \leq n -m + l_{min}\\ 
 X_{11_j} &  \text{if } n - m + l_{min} < j \leq n
\end{array} \right.
\end{equation*}

\begin{figure}[htb]
 \begin{center}

  \resizebox{3.95in}{!}{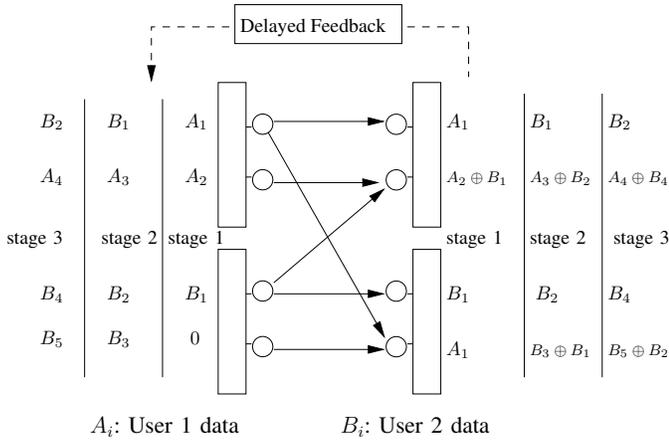}
   \caption {Three stages of one link feedback; $\frac{n}{m} = \frac{1}{2}$}
  \label{fig:one-sided feedback}
  \end{center}
\end{figure}

\noindent \emph{Feedback slot 1:} The received vector at receiver 1 is fed back to the $1^{\mathrm{st}}$ transmitter. The transmitter is capable of resolving the data with interference.

\noindent \emph{Time slot $i$:} The transmitted vector at the $i^{\mathrm{th}}$ time slot are:

\begin{equation*}
 X_{1i_j} = \left\{
\begin{array}{ll}
X_{2(i-1)_j} & \text{if }  1 \leq j \leq l_{min } \\
 \text{i.i.d. data}   &  \text{if } l_{min} < j \leq n 

\end{array} \right.
\end{equation*}
where the lower ($ n - \min(m,n-m)$ ) bits are independent data bits. The user 2 transmits data vector such that the lower $(2m - n)^{+}$ bits are set to 0 while the rest of the bits are all independent. The received vector $Y_{1i}$ and $Y_{2i}$ at receiver 1 and 2 respectively are

\begin{equation*}
  Y_{1i_j} = \left\{
\begin{array}{ll}
X_{2({i-1})_j} & \text{if }  1 \leq j \leq l_{min} \\
 X_{1i_j}     & \text{if }  l_{min} < j \leq l_{max}   \\
X_{1i_j} \oplus X_{2i_{j - l_{max}}} &  \text{if }  l_{max} < j \leq n

\end{array} \right.
\end{equation*}

\begin{equation*}
  Y_{2i_j} = \left\{
\begin{array}{ll}
X_{2i_j} & \text{if }  1 \leq j \leq n-m \\
 X_{2i_j} \oplus X_{2(i-1)_{j - n + m}}   &  \text{if }  n - m  < j  \\ & \leq  \min(n, 2(n- m))   \\
X_{2i_{j - \min(n, 2(n-m))}} \ & \text{if } \min(n, 2(n-m)) < j \\ &  \leq n

\end{array} \right.
\end{equation*}

We observe here that at the first receiver $\max(n -m ,m)$ bits are decoded in every time slot. Also among these $\min(n-m,m)$ bits are the undecoded bits of $(i -1)^{\mathrm{th}}$ time slot, whereas the same number of bits of the $i^{\mathrm{th}}$ time slot are decoded in the next time slot. All the received bits at the receiver 2 are decoded. The sum total off data bits, $B_{\mathrm{sum}}$ transmitted after $T$ time slots and $T-1$ feedback slots is
\begin{equation}
\begin{array}{lcl}
 B_{\mathrm{sum}}  & = &  (T-1)\times\left(\max(n - m,m) + \min(n,2(n -m))\right)  \\ &  + &  \max(n -m,m) + (n-m)
\end{array}
\end{equation}

 The average sum capacity for one-sided feedback, $C_{sum}$ is therefore lower bounded by
\begin{equation}
  C_{sum}\geq \mathop{\lim}\limits_{T \to \infty} (1 - \frac{1}{T})\times \left(\max(n - m,m) + \min(n,2(n-m) )\right)
\end{equation}

\noindent Observe that when $\frac{m}{n} \leq \frac{1}{2}$ the expression can be rewritten as
\begin{equation}
  C_{sum} \geq \mathop{\lim}\limits_{T \to \infty} (1 - \frac{1}{T})\times(n - m + n)
\end{equation}
In the regime where $\frac{m}{n} \geq \frac{1}{2}$, the expression is equivalent to
\begin{equation}
  C_{sum} \geq \mathop{\lim}\limits_{T \to \infty} (1 - \frac{1}{T})\times\left(m + 2(n-m)\right)
\end{equation}
Thus, in either case the sum-capacity can be shown to approaching ($2n - m$), which is the upper bound on symmetric feedback sum rate capacity.

\noindent \emph{Case 2, $2/3 \le m/n\le 2$}: Two link feedback do not increase sum capacity and hence one link feedback can achieve the same sum capacity.

\noindent \emph{Case 3, $m/n\ge 2$}: In this case, we will show that the $(R_1,R_2)=(0,m)$ can be achieved which achieves the optimal sum capacity. To see this, consider that the second user sends data on all the links. It receives $n$ bits directly from the transmitter while the remaining $m-n$ bits are routed to the first receiver which sends it to the first transmitter and are then passed without interference to the second receiver. Hence, the total number of noiseless paths from the second transmitter to the second receiver are $m$ thus resulting in a sum feedback capacity of $m$.

\end{section}

\end{document}

%% file: symmdeter.pdf_t
\begin{picture}(0,0)%
\includegraphics{symmdeter.pdf}%
\end{picture}%
\setlength{\unitlength}{4185sp}%
\begingroup\makeatletter\ifx\SetFigFontNFSS\undefined%
\gdef\SetFigFontNFSS#1#2#3#4#5{%
  \reset@font\fontsize{#1}{#2pt}%
  \fontfamily{#3}\fontseries{#4}\fontshape{#5}%
  \selectfont}%
\fi\endgroup%
\begin{picture}(2865,1935)(2731,-2344)
\put(3286,-2266){\makebox(0,0)[lb]{\smash{{\SetFigFontNFSS{10}{12.0}{\familydefault}{\mddefault}{\updefault}{\color[rgb]{0,0,0}$n_{22} = n$}%
}}}}
\put(3196,-1006){\makebox(0,0)[lb]{\smash{{\SetFigFontNFSS{12}{14.4}{\familydefault}{\mddefault}{\updefault}{\color[rgb]{0,0,0}Tx$_1$}%
}}}}
\put(3196,-1906){\makebox(0,0)[lb]{\smash{{\SetFigFontNFSS{12}{14.4}{\familydefault}{\mddefault}{\updefault}{\color[rgb]{0,0,0}Tx$_2$}%
}}}}
\put(5581,-1906){\makebox(0,0)[lb]{\smash{{\SetFigFontNFSS{12}{14.4}{\familydefault}{\mddefault}{\updefault}{\color[rgb]{0,0,0}$Y_2$}%
}}}}
\put(5581,-1006){\makebox(0,0)[lb]{\smash{{\SetFigFontNFSS{12}{14.4}{\familydefault}{\mddefault}{\updefault}{\color[rgb]{0,0,0}$Y_1$}%
}}}}
\put(5041,-1006){\makebox(0,0)[lb]{\smash{{\SetFigFontNFSS{12}{14.4}{\familydefault}{\mddefault}{\updefault}{\color[rgb]{0,0,0}Rx$_1$}%
}}}}
\put(5041,-1906){\makebox(0,0)[lb]{\smash{{\SetFigFontNFSS{12}{14.4}{\familydefault}{\mddefault}{\updefault}{\color[rgb]{0,0,0}Rx$_2$}%
}}}}
\put(4636,-1366){\makebox(0,0)[lb]{\smash{{\SetFigFontNFSS{12}{14.4}{\familydefault}{\mddefault}{\updefault}{\color[rgb]{0,0,0}$V_2$}%
}}}}
\put(4591,-2266){\makebox(0,0)[lb]{\smash{{\SetFigFontNFSS{12}{14.4}{\familydefault}{\mddefault}{\updefault}{\color[rgb]{0,0,0}$V_1$}%
}}}}
\put(2746,-1051){\makebox(0,0)[lb]{\smash{{\SetFigFontNFSS{12}{14.4}{\familydefault}{\mddefault}{\updefault}{\color[rgb]{0,0,0}$X_1$}%
}}}}
\put(2746,-1951){\makebox(0,0)[lb]{\smash{{\SetFigFontNFSS{12}{14.4}{\familydefault}{\mddefault}{\updefault}{\color[rgb]{0,0,0}$X_2$}%
}}}}
\put(3826,-556){\makebox(0,0)[lb]{\smash{{\SetFigFontNFSS{10}{12.0}{\familydefault}{\mddefault}{\updefault}{\color[rgb]{0,0,0}$n_{12} = m$}%
}}}}
\put(3556,-1546){\makebox(0,0)[lb]{\smash{{\SetFigFontNFSS{10}{12.0}{\familydefault}{\mddefault}{\updefault}{\color[rgb]{0,0,0}$n_{21} = m$}%
}}}}
\put(3286,-1321){\makebox(0,0)[lb]{\smash{{\SetFigFontNFSS{10}{12.0}{\familydefault}{\mddefault}{\updefault}{\color[rgb]{0,0,0}$n_{11} = n$}%
}}}}
\end{picture}%

%% file: onesided_model.pdf_t
\begin{picture}(0,0)%
\includegraphics{onesided_model.pdf}%
\end{picture}%
\setlength{\unitlength}{4185sp}%
\begingroup\makeatletter\ifx\SetFigFontNFSS\undefined%
\gdef\SetFigFontNFSS#1#2#3#4#5{%
  \reset@font\fontsize{#1}{#2pt}%
  \fontfamily{#3}\fontseries{#4}\fontshape{#5}%
  \selectfont}%
\fi\endgroup%
\begin{picture}(3087,2682)(2731,-2686)
\put(3826,-1501){\makebox(0,0)[lb]{\smash{{\SetFigFontNFSS{10}{12.0}{\familydefault}{\mddefault}{\updefault}{\color[rgb]{0,0,0}$m$}%
}}}}
\put(3196,-1006){\makebox(0,0)[lb]{\smash{{\SetFigFontNFSS{12}{14.4}{\familydefault}{\mddefault}{\updefault}{\color[rgb]{0,0,0}Tx$_1$}%
}}}}
\put(3196,-1906){\makebox(0,0)[lb]{\smash{{\SetFigFontNFSS{12}{14.4}{\familydefault}{\mddefault}{\updefault}{\color[rgb]{0,0,0}Tx$_2$}%
}}}}
\put(5581,-1906){\makebox(0,0)[lb]{\smash{{\SetFigFontNFSS{12}{14.4}{\familydefault}{\mddefault}{\updefault}{\color[rgb]{0,0,0}$Y_2$}%
}}}}
\put(5581,-1006){\makebox(0,0)[lb]{\smash{{\SetFigFontNFSS{12}{14.4}{\familydefault}{\mddefault}{\updefault}{\color[rgb]{0,0,0}$Y_1$}%
}}}}
\put(5041,-1006){\makebox(0,0)[lb]{\smash{{\SetFigFontNFSS{12}{14.4}{\familydefault}{\mddefault}{\updefault}{\color[rgb]{0,0,0}Rx$_1$}%
}}}}
\put(5041,-1906){\makebox(0,0)[lb]{\smash{{\SetFigFontNFSS{12}{14.4}{\familydefault}{\mddefault}{\updefault}{\color[rgb]{0,0,0}Rx$_2$}%
}}}}
\put(4636,-1366){\makebox(0,0)[lb]{\smash{{\SetFigFontNFSS{12}{14.4}{\familydefault}{\mddefault}{\updefault}{\color[rgb]{0,0,0}$V_2$}%
}}}}
\put(4591,-2266){\makebox(0,0)[lb]{\smash{{\SetFigFontNFSS{12}{14.4}{\familydefault}{\mddefault}{\updefault}{\color[rgb]{0,0,0}$V_1$}%
}}}}
\put(2746,-1051){\makebox(0,0)[lb]{\smash{{\SetFigFontNFSS{12}{14.4}{\familydefault}{\mddefault}{\updefault}{\color[rgb]{0,0,0}$X_1$}%
}}}}
\put(2746,-1951){\makebox(0,0)[lb]{\smash{{\SetFigFontNFSS{12}{14.4}{\familydefault}{\mddefault}{\updefault}{\color[rgb]{0,0,0}$X_2$}%
}}}}
\put(3556,-2221){\makebox(0,0)[lb]{\smash{{\SetFigFontNFSS{10}{12.0}{\familydefault}{\mddefault}{\updefault}{\color[rgb]{0,0,0}$n$}%
}}}}
\put(3511,-1321){\makebox(0,0)[lb]{\smash{{\SetFigFontNFSS{10}{12.0}{\familydefault}{\mddefault}{\updefault}{\color[rgb]{0,0,0}$n$}%
}}}}
\put(3871,-646){\makebox(0,0)[lb]{\smash{{\SetFigFontNFSS{10}{12.0}{\familydefault}{\mddefault}{\updefault}{\color[rgb]{0,0,0}$m$}%
}}}}
\end{picture}%

%% file: fourlink.pdf_t
\begin{picture}(0,0)%
\includegraphics{fourlink.pdf}%
\end{picture}%
\setlength{\unitlength}{4185sp}%
\begingroup\makeatletter\ifx\SetFigFontNFSS\undefined%
\gdef\SetFigFontNFSS#1#2#3#4#5{%
  \reset@font\fontsize{#1}{#2pt}%
  \fontfamily{#3}\fontseries{#4}\fontshape{#5}%
  \selectfont}%
\fi\endgroup%
\begin{picture}(3132,3399)(2731,-3403)
\put(3196,-1006){\makebox(0,0)[lb]{\smash{{\SetFigFontNFSS{12}{14.4}{\familydefault}{\mddefault}{\updefault}{\color[rgb]{0,0,0}Tx$_1$}%
}}}}
\put(5581,-1006){\makebox(0,0)[lb]{\smash{{\SetFigFontNFSS{12}{14.4}{\familydefault}{\mddefault}{\updefault}{\color[rgb]{0,0,0}$Y_1$}%
}}}}
\put(5041,-1006){\makebox(0,0)[lb]{\smash{{\SetFigFontNFSS{12}{14.4}{\familydefault}{\mddefault}{\updefault}{\color[rgb]{0,0,0}Rx$_1$}%
}}}}
\put(4636,-1366){\makebox(0,0)[lb]{\smash{{\SetFigFontNFSS{12}{14.4}{\familydefault}{\mddefault}{\updefault}{\color[rgb]{0,0,0}$V_2$}%
}}}}
\put(2746,-1051){\makebox(0,0)[lb]{\smash{{\SetFigFontNFSS{12}{14.4}{\familydefault}{\mddefault}{\updefault}{\color[rgb]{0,0,0}$X_1$}%
}}}}
\put(5581,-2536){\makebox(0,0)[lb]{\smash{{\SetFigFontNFSS{12}{14.4}{\familydefault}{\mddefault}{\updefault}{\color[rgb]{0,0,0}$Y_2$}%
}}}}
\put(5041,-2536){\makebox(0,0)[lb]{\smash{{\SetFigFontNFSS{12}{14.4}{\familydefault}{\mddefault}{\updefault}{\color[rgb]{0,0,0}Rx$_2$}%
}}}}
\put(4591,-2896){\makebox(0,0)[lb]{\smash{{\SetFigFontNFSS{12}{14.4}{\familydefault}{\mddefault}{\updefault}{\color[rgb]{0,0,0}$V_1$}%
}}}}
\put(3196,-2536){\makebox(0,0)[lb]{\smash{{\SetFigFontNFSS{12}{14.4}{\familydefault}{\mddefault}{\updefault}{\color[rgb]{0,0,0}Tx$_2$}%
}}}}
\put(2746,-2581){\makebox(0,0)[lb]{\smash{{\SetFigFontNFSS{12}{14.4}{\familydefault}{\mddefault}{\updefault}{\color[rgb]{0,0,0}$X_2$}%
}}}}
\put(3871,-691){\makebox(0,0)[lb]{\smash{{\SetFigFontNFSS{10}{12.0}{\familydefault}{\mddefault}{\updefault}{\color[rgb]{0,0,0}$m$}%
}}}}
\put(4051,-2311){\makebox(0,0)[lb]{\smash{{\SetFigFontNFSS{10}{12.0}{\familydefault}{\mddefault}{\updefault}{\color[rgb]{0,0,0}$m$}%
}}}}
\put(3511,-1321){\makebox(0,0)[lb]{\smash{{\SetFigFontNFSS{10}{12.0}{\familydefault}{\mddefault}{\updefault}{\color[rgb]{0,0,0}$n$}%
}}}}
\put(3511,-2851){\makebox(0,0)[lb]{\smash{{\SetFigFontNFSS{10}{12.0}{\familydefault}{\mddefault}{\updefault}{\color[rgb]{0,0,0}$n$}%
}}}}
\put(3826,-196){\makebox(0,0)[lb]{\smash{{\SetFigFontNFSS{8}{9.6}{\familydefault}{\mddefault}{\updefault}{\color[rgb]{0,0,0}Delayed Feedback}%
}}}}
\put(3826,-3301){\makebox(0,0)[lb]{\smash{{\SetFigFontNFSS{8}{9.6}{\familydefault}{\mddefault}{\updefault}{\color[rgb]{0,0,0}Delayed Feedback}%
}}}}
\put(4591,-1636){\makebox(0,0)[lb]{\smash{{\SetFigFontNFSS{8}{9.6}{\familydefault}{\mddefault}{\updefault}{\color[rgb]{0,0,0}Delayed Feedback}%
}}}}
\put(4591,-1906){\makebox(0,0)[lb]{\smash{{\SetFigFontNFSS{8}{9.6}{\familydefault}{\mddefault}{\updefault}{\color[rgb]{0,0,0}Delayed Feedback}%
}}}}
\end{picture}%

%% file: relay.pdf_t
\begin{picture}(0,0)%
\includegraphics{relay.pdf}%
\end{picture}%
\setlength{\unitlength}{4144sp}%
\begingroup\makeatletter\ifx\SetFigFontNFSS\undefined%
\gdef\SetFigFontNFSS#1#2#3#4#5{%
  \reset@font\fontsize{#1}{#2pt}%
  \fontfamily{#3}\fontseries{#4}\fontshape{#5}%
  \selectfont}%
\fi\endgroup%
\begin{picture}(3174,1824)(2599,-3178)
\put(2701,-2941){\makebox(0,0)[lb]{\smash{{\SetFigFontNFSS{12}{14.4}{\familydefault}{\mddefault}{\updefault}{\color[rgb]{0,0,0}$\mathrm{Tx}_2$}%
}}}}
\put(5266,-2941){\makebox(0,0)[lb]{\smash{{\SetFigFontNFSS{12}{14.4}{\familydefault}{\mddefault}{\updefault}{\color[rgb]{0,0,0}$\mathrm{Rx}_2$}%
}}}}
\put(4051,-1816){\makebox(0,0)[lb]{\smash{{\SetFigFontNFSS{12}{14.4}{\familydefault}{\mddefault}{\updefault}{\color[rgb]{0,0,0}$\mathrm{Rx}_1$}%
}}}}
\put(3331,-2266){\makebox(0,0)[lb]{\smash{{\SetFigFontNFSS{12}{14.4}{\familydefault}{\mddefault}{\updefault}{\color[rgb]{0,0,0}$m$}%
}}}}
\put(4051,-1591){\makebox(0,0)[lb]{\smash{{\SetFigFontNFSS{12}{14.4}{\familydefault}{\mddefault}{\updefault}{\color[rgb]{0,0,0}$\mathrm{Tx}_1$}%
}}}}
\put(4816,-2176){\makebox(0,0)[lb]{\smash{{\SetFigFontNFSS{12}{14.4}{\familydefault}{\mddefault}{\updefault}{\color[rgb]{0,0,0}$m$}%
}}}}
\put(4051,-2761){\makebox(0,0)[lb]{\smash{{\SetFigFontNFSS{12}{14.4}{\familydefault}{\mddefault}{\updefault}{\color[rgb]{0,0,0}$n$}%
}}}}
\end{picture}%

%% file: reciproc.pdf_t
\begin{picture}(0,0)%
\includegraphics{reciproc.pdf}%
\end{picture}%
\setlength{\unitlength}{4185sp}%
\begingroup\makeatletter\ifx\SetFigFontNFSS\undefined%
\gdef\SetFigFontNFSS#1#2#3#4#5{%
  \reset@font\fontsize{#1}{#2pt}%
  \fontfamily{#3}\fontseries{#4}\fontshape{#5}%
  \selectfont}%
\fi\endgroup%
\begin{picture}(2409,1843)(3139,-2276)
\put(3826,-556){\makebox(0,0)[lb]{\smash{{\SetFigFontNFSS{10}{12.0}{\familydefault}{\mddefault}{\updefault}{\color[rgb]{0,0,0}$m$}%
}}}}
\put(3826,-1501){\makebox(0,0)[lb]{\smash{{\SetFigFontNFSS{10}{12.0}{\familydefault}{\mddefault}{\updefault}{\color[rgb]{0,0,0}$m$}%
}}}}
\put(3196,-1006){\makebox(0,0)[lb]{\smash{{\SetFigFontNFSS{12}{14.4}{\familydefault}{\mddefault}{\updefault}{\color[rgb]{0,0,0}Rx$_1$}%
}}}}
\put(3196,-1906){\makebox(0,0)[lb]{\smash{{\SetFigFontNFSS{12}{14.4}{\familydefault}{\mddefault}{\updefault}{\color[rgb]{0,0,0}Rx$_2$}%
}}}}
\put(5041,-1006){\makebox(0,0)[lb]{\smash{{\SetFigFontNFSS{12}{14.4}{\familydefault}{\mddefault}{\updefault}{\color[rgb]{0,0,0}Tx$_1$}%
}}}}
\put(5041,-1906){\makebox(0,0)[lb]{\smash{{\SetFigFontNFSS{12}{14.4}{\familydefault}{\mddefault}{\updefault}{\color[rgb]{0,0,0}Tx$_2$}%
}}}}
\put(3511,-2221){\makebox(0,0)[lb]{\smash{{\SetFigFontNFSS{10}{12.0}{\familydefault}{\mddefault}{\updefault}{\color[rgb]{0,0,0}$n$}%
}}}}
\put(3511,-1321){\makebox(0,0)[lb]{\smash{{\SetFigFontNFSS{10}{12.0}{\familydefault}{\mddefault}{\updefault}{\color[rgb]{0,0,0}$n$}%
}}}}
\end{picture}%

%% file: halfdp.pdf_t
\begin{picture}(0,0)%
\includegraphics{halfdp.pdf}%
\end{picture}%
\setlength{\unitlength}{4144sp}%
\begingroup\makeatletter\ifx\SetFigFontNFSS\undefined%
\gdef\SetFigFontNFSS#1#2#3#4#5{%
  \reset@font\fontsize{#1}{#2pt}%
  \fontfamily{#3}\fontseries{#4}\fontshape{#5}%
  \selectfont}%
\fi\endgroup%
\begin{picture}(2814,3672)(2689,-4123)
\put(2836,-1681){\makebox(0,0)[lb]{\smash{{\SetFigFontNFSS{14}{16.8}{\familydefault}{\mddefault}{\updefault}{\color[rgb]{0,0,0}$\mathrm{Tx}_1$}%
}}}}
\put(2836,-3256){\makebox(0,0)[lb]{\smash{{\SetFigFontNFSS{14}{16.8}{\familydefault}{\mddefault}{\updefault}{\color[rgb]{0,0,0}$\mathrm{Tx}_2$}%
}}}}
\put(3871,-4021){\makebox(0,0)[lb]{\smash{{\SetFigFontNFSS{14}{16.8}{\familydefault}{\mddefault}{\updefault}{\color[rgb]{0,0,0}$n(1 -t)$}%
}}}}
\put(3871,-1591){\makebox(0,0)[lb]{\smash{{\SetFigFontNFSS{14}{16.8}{\familydefault}{\mddefault}{\updefault}{\color[rgb]{0,0,0}$nt$}%
}}}}
\put(3961,-3076){\makebox(0,0)[lb]{\smash{{\SetFigFontNFSS{14}{16.8}{\familydefault}{\mddefault}{\updefault}{\color[rgb]{0,0,0}$nt$}%
}}}}
\put(3781,-646){\makebox(0,0)[lb]{\smash{{\SetFigFontNFSS{14}{16.8}{\familydefault}{\mddefault}{\updefault}{\color[rgb]{0,0,0}$n(1-t)$}%
}}}}
\put(4951,-3256){\makebox(0,0)[lb]{\smash{{\SetFigFontNFSS{14}{16.8}{\familydefault}{\mddefault}{\updefault}{\color[rgb]{0,0,0}$\mathrm{Rx}_2$}%
}}}}
\put(4951,-1681){\makebox(0,0)[lb]{\smash{{\SetFigFontNFSS{14}{16.8}{\familydefault}{\mddefault}{\updefault}{\color[rgb]{0,0,0}$\mathrm{Rx}_1$}%
}}}}
\end{picture}%

%% file: halfdp2.pdf_t
\begin{picture}(0,0)%
\includegraphics{halfdp2.pdf}%
\end{picture}%
\setlength{\unitlength}{4144sp}%
\begingroup\makeatletter\ifx\SetFigFontNFSS\undefined%
\gdef\SetFigFontNFSS#1#2#3#4#5{%
  \reset@font\fontsize{#1}{#2pt}%
  \fontfamily{#3}\fontseries{#4}\fontshape{#5}%
  \selectfont}%
\fi\endgroup%
\begin{picture}(2274,972)(6289,-2863)
\put(7966,-2266){\makebox(0,0)[lb]{\smash{{\SetFigFontNFSS{14}{16.8}{\familydefault}{\mddefault}{\updefault}{\color[rgb]{0,0,0}$\mathrm{Tx}_2$}%
}}}}
\put(7246,-2491){\makebox(0,0)[lb]{\smash{{\SetFigFontNFSS{14}{16.8}{\familydefault}{\mddefault}{\updefault}{\color[rgb]{0,0,0}$nt$}%
}}}}
\put(6391,-2266){\makebox(0,0)[lb]{\smash{{\SetFigFontNFSS{14}{16.8}{\familydefault}{\mddefault}{\updefault}{\color[rgb]{0,0,0}$\mathrm{Tx}_1$}%
}}}}
\put(6391,-2536){\makebox(0,0)[lb]{\smash{{\SetFigFontNFSS{14}{16.8}{\familydefault}{\mddefault}{\updefault}{\color[rgb]{0,0,0}$\mathrm{Rx}_2$}%
}}}}
\put(7111,-2086){\makebox(0,0)[lb]{\smash{{\SetFigFontNFSS{14}{16.8}{\familydefault}{\mddefault}{\updefault}{\color[rgb]{0,0,0}$n(1-t)$}%
}}}}
\put(7966,-2521){\makebox(0,0)[lb]{\smash{{\SetFigFontNFSS{14}{16.8}{\familydefault}{\mddefault}{\updefault}{\color[rgb]{0,0,0}$\mathrm{Rx}_1$}%
}}}}
\end{picture}%

%% file: onesided_ab.pdf_t
\begin{picture}(0,0)%
\includegraphics{onesided_ab.pdf}%
\end{picture}%
\setlength{\unitlength}{4185sp}%
\begingroup\makeatletter\ifx\SetFigFontNFSS\undefined%
\gdef\SetFigFontNFSS#1#2#3#4#5{%
  \reset@font\fontsize{#1}{#2pt}%
  \fontfamily{#3}\fontseries{#4}\fontshape{#5}%
  \selectfont}%
\fi\endgroup%
\begin{picture}(6102,3555)(1426,-3199)
\put(2251,-646){\makebox(0,0)[lb]{\smash{{\SetFigFontNFSS{10}{12.0}{\familydefault}{\mddefault}{\updefault}{\color[rgb]{0,0,0}$B_1$}%
}}}}
\put(2251,-1096){\makebox(0,0)[lb]{\smash{{\SetFigFontNFSS{10}{12.0}{\familydefault}{\mddefault}{\updefault}{\color[rgb]{0,0,0}$A_3$}%
}}}}
\put(2251,-2041){\makebox(0,0)[lb]{\smash{{\SetFigFontNFSS{10}{12.0}{\familydefault}{\mddefault}{\updefault}{\color[rgb]{0,0,0}$B_2$}%
}}}}
\put(2251,-2401){\makebox(0,0)[lb]{\smash{{\SetFigFontNFSS{10}{12.0}{\familydefault}{\mddefault}{\updefault}{\color[rgb]{0,0,0}$B_3$}%
}}}}
\put(1711,-646){\makebox(0,0)[lb]{\smash{{\SetFigFontNFSS{10}{12.0}{\familydefault}{\mddefault}{\updefault}{\color[rgb]{0,0,0}$B_2$}%
}}}}
\put(1711,-1096){\makebox(0,0)[lb]{\smash{{\SetFigFontNFSS{10}{12.0}{\familydefault}{\mddefault}{\updefault}{\color[rgb]{0,0,0}$A_4$}%
}}}}
\put(1711,-2041){\makebox(0,0)[lb]{\smash{{\SetFigFontNFSS{10}{12.0}{\familydefault}{\mddefault}{\updefault}{\color[rgb]{0,0,0}$B_4$}%
}}}}
\put(1711,-2401){\makebox(0,0)[lb]{\smash{{\SetFigFontNFSS{10}{12.0}{\familydefault}{\mddefault}{\updefault}{\color[rgb]{0,0,0}$B_5$}%
}}}}
\put(4996,-646){\makebox(0,0)[lb]{\smash{{\SetFigFontNFSS{10}{12.0}{\familydefault}{\mddefault}{\updefault}{\color[rgb]{0,0,0}$A_1$}%
}}}}
\put(4996,-2041){\makebox(0,0)[lb]{\smash{{\SetFigFontNFSS{10}{12.0}{\familydefault}{\mddefault}{\updefault}{\color[rgb]{0,0,0}$B_1$}%
}}}}
\put(4996,-2491){\makebox(0,0)[lb]{\smash{{\SetFigFontNFSS{10}{12.0}{\familydefault}{\mddefault}{\updefault}{\color[rgb]{0,0,0}$A_1$}%
}}}}
\put(2881,-646){\makebox(0,0)[lb]{\smash{{\SetFigFontNFSS{10}{12.0}{\familydefault}{\mddefault}{\updefault}{\color[rgb]{0,0,0}$A_1$}%
}}}}
\put(2881,-1096){\makebox(0,0)[lb]{\smash{{\SetFigFontNFSS{10}{12.0}{\familydefault}{\mddefault}{\updefault}{\color[rgb]{0,0,0}$A_2$}%
}}}}
\put(2881,-2041){\makebox(0,0)[lb]{\smash{{\SetFigFontNFSS{10}{12.0}{\familydefault}{\mddefault}{\updefault}{\color[rgb]{0,0,0}$B_1$}%
}}}}
\put(2926,-2401){\makebox(0,0)[lb]{\smash{{\SetFigFontNFSS{10}{12.0}{\familydefault}{\mddefault}{\updefault}{\color[rgb]{0,0,0}$0$}%
}}}}
\put(4996,-1096){\makebox(0,0)[lb]{\smash{{\SetFigFontNFSS{8}{9.6}{\familydefault}{\mddefault}{\updefault}{\color[rgb]{0,0,0}$A_2 \oplus B_1$}%
}}}}
\put(4996,-1591){\makebox(0,0)[lb]{\smash{{\SetFigFontNFSS{10}{12.0}{\familydefault}{\mddefault}{\updefault}{\color[rgb]{0,0,0}$\text{stage 1}$}%
}}}}
\put(2746,-1591){\makebox(0,0)[lb]{\smash{{\SetFigFontNFSS{10}{12.0}{\familydefault}{\mddefault}{\updefault}{\color[rgb]{0,0,0}$\text{stage 1}$}%
}}}}
\put(2206,-1591){\makebox(0,0)[lb]{\smash{{\SetFigFontNFSS{10}{12.0}{\familydefault}{\mddefault}{\updefault}{\color[rgb]{0,0,0}$\text{stage 2}$}%
}}}}
\put(5671,-1591){\makebox(0,0)[lb]{\smash{{\SetFigFontNFSS{10}{12.0}{\familydefault}{\mddefault}{\updefault}{\color[rgb]{0,0,0}$\text{stage 2}$}%
}}}}
\put(6346,-1591){\makebox(0,0)[lb]{\smash{{\SetFigFontNFSS{10}{12.0}{\familydefault}{\mddefault}{\updefault}{\color[rgb]{0,0,0}$\text{stage 3}$}%
}}}}
\put(6301,-646){\makebox(0,0)[lb]{\smash{{\SetFigFontNFSS{10}{12.0}{\familydefault}{\mddefault}{\updefault}{\color[rgb]{0,0,0}$B_2$}%
}}}}
\put(6301,-1096){\makebox(0,0)[lb]{\smash{{\SetFigFontNFSS{8}{9.6}{\familydefault}{\mddefault}{\updefault}{\color[rgb]{0,0,0}$A_4 \oplus B_4$}%
}}}}
\put(6301,-2491){\makebox(0,0)[lb]{\smash{{\SetFigFontNFSS{8}{9.6}{\familydefault}{\mddefault}{\updefault}{\color[rgb]{0,0,0}$B_5\oplus B_2$}%
}}}}
\put(6301,-2041){\makebox(0,0)[lb]{\smash{{\SetFigFontNFSS{10}{12.0}{\familydefault}{\mddefault}{\updefault}{\color[rgb]{0,0,0}$B_4$}%
}}}}
\put(5716,-2041){\makebox(0,0)[lb]{\smash{{\SetFigFontNFSS{10}{12.0}{\familydefault}{\mddefault}{\updefault}{\color[rgb]{0,0,0}$B_2$}%
}}}}
\put(5671,-646){\makebox(0,0)[lb]{\smash{{\SetFigFontNFSS{10}{12.0}{\familydefault}{\mddefault}{\updefault}{\color[rgb]{0,0,0}$B_1$}%
}}}}
\put(5671,-1096){\makebox(0,0)[lb]{\smash{{\SetFigFontNFSS{8}{9.6}{\familydefault}{\mddefault}{\updefault}{\color[rgb]{0,0,0}$A_3 \oplus B_2$}%
}}}}
\put(5671,-2491){\makebox(0,0)[lb]{\smash{{\SetFigFontNFSS{8}{9.6}{\familydefault}{\mddefault}{\updefault}{\color[rgb]{0,0,0}$B_3\oplus B_1$}%
}}}}
\put(2116,-3121){\makebox(0,0)[lb]{\smash{{\SetFigFontNFSS{12}{14.4}{\familydefault}{\mddefault}{\updefault}{\color[rgb]{0,0,0}$A_i$: User 1 data}%
}}}}
\put(4141,-3121){\makebox(0,0)[lb]{\smash{{\SetFigFontNFSS{12}{14.4}{\familydefault}{\mddefault}{\updefault}{\color[rgb]{0,0,0}$B_i$: User 2 data }%
}}}}
\put(1441,-1591){\makebox(0,0)[lb]{\smash{{\SetFigFontNFSS{10}{12.0}{\familydefault}{\mddefault}{\updefault}{\color[rgb]{0,0,0}$\text{stage 3}$}%
}}}}
\end{picture}%